\documentclass[journal]{IEEEtran}
\usepackage{graphicx}
\usepackage{amsmath}
\usepackage{algorithmic}
\usepackage{subfigure}
\usepackage{multirow}
\usepackage{array}

\begin{document}

\title{On Link Availability Probability of Routing Protocols for Urban Scenario in VANETs}

\author{\IEEEauthorblockN{S. Kumar, N. Javaid, Z. Yousuf, H. Kumar, Z. A. Khan$^{\$}$, U. Qasim$^{\ddag}$\\\vspace{0.4cm}}

        COMSATS Institute of Information Technology, Islamabad, Pakistan. \\
        $^{\$}$Faculty of Engineering, Dalhousie University, Halifax, Canada.\\
                        $^{\ddag}$University of Alberta, Alberta, Canada.

     }

%

\maketitle

\begin{abstract}
This paper presents the link availability probability. We evaluate and compare the link availability probability for routing protocols; Ad hoc On-demand Distance vector (AODV), Dynamic Source Routing (DSR) and Fisheye State Routing (FSR) for different number of connections and node density. A novel contribution of this work is enhancement in existing parameters of routing protocols; AODV, DSR and FSR as MOD-AODV, MOD-DSR and MOD-FSR. From the results, we observe that MOD-DSR and DSR outperform MOD-AODV, AODV, MOD-OLSR and OLSR in terms of Packet Delivery Ratio (PDR), Average End-to End Delay (AE2ED), link availability probability at the cost of high value of Normalized Routing Overhead (NRO).
\end{abstract}

\IEEEpeerreviewmaketitle
\section{Introduction}
In VANETs, any permanent infrastructure is not necessarily required and vehicles may communicate with each other through self-organized network within the limited range of few hundred meters. VANETs may rightly be said a sub network of MANETs. With high mobility due to the high motion of vehicles, its topology rapidly changes. To make realize the vehicles what is happening around them, to reduce the number of road accidents and thereby increasing road safety, vehicles should communicate with each other. It is of great importance that information transmitted should be sufficient for safety and without latency. The main features of VANETs are to provide safety to vehicles, to avoid accidents or collisions, awareness of traffic jams, access of internet for passengers and multimedia entertainment services.

In this paper, we evaluate the performance of three routing protocols; AODV [1], DSR [2] and FSR [3] for VANETs in urban scenario. The evaluation of both default and modified routing protocols has done with performance parameter; PDR, AE2ED and NRO with varying node densities and different number of connections. In case of urban scenario for communication, vehicles can move in any direction. In other words, vehicles can move in both directions (same and opposite). So, we find out link availability probability between nodes with different cases that is described in link availability probability section.


\section{Related Work}

In last few years, efforts are made on work that relate to our work. In [4], authors evaluate radio propagation models. Through the study, it is observed that Nakagami model is able to give accurate simulation results for VANETs in urban scenario.

Studying in [5], authors improve the routing performance of routing protocols in high mobility and high density VANETs. They propose two different cases of vehicles velocity and find the link available probability for these two cases. For further improvements in PDR and AE2ED, they present two algorithms.

[6] discusses the comparison and the performance of routing protocols; AODV, DSDV and DSR in VANETs, with the extensive simulation studies for highway scenarios. Simulation results show the performance parameters; PDR, AE2ED and NRL, and concluded that the routing protocols are unsuitable for VANETs, but dedicated for MANETs.

Another study in [7], evaluates the performance of routing protocols; AODV, OLSR and Modified OLSR, under realistic radio channel characteristic using NS-2 Nakagami fading model. They analyze the performance of both routing protocols with performance parameters; PDR, AE2ED and NRL with varying node densities and different number of connections.

\section{Motivation}
In this paper we have done simulation in urban scenario that was due to motivated by [4],[5] and [7].

In [4], they have done evaluation of protocols through radio propagation model Nakagami by the knowledge as the best model among all these models. Authors in [5] have analyzed the problem for two different cases of velocity of the nodes and also proposed two algorithm for improving routing performance in high scalability and high mobility. In [7], authors evaluate the routing protocols; AODV, OLSR and Modified OLSR in VANETs urban scenario.

Hence in [4], we study that Nakagami model is best among all the radio propagation models so for our simulation work, we used this model. As inspired by [5] and [7], we evaluate and compare the performance of both default and modified routing protocols; AODV, DSR and FSR in urban scenario with more different cases of velocity between nodes. Further we also determine the link availability probability between nodes.

\section{Link Availability Probability}
In [5], authors discuss the mobility scenario in VANETs taking two cases of velocity; (i) when both nodes have same velocity and (ii) when both have different velocities. After taking assumption they find out the relative velocity and expected relative velocity between nodes. They determine the link availability probability for these two cases using simple area of covered region between two nodes and radio covering range of any node.

We consider an urban scenario in VANETs in which nodes (vehicles) can move in both the directions (same and opposite). Assumptions are taken that two nodes are moving with velocities $\vec {v_1}$ and $\vec {v_{2}}$ respectively, the distance between two nodes is \emph{\textbf{d}} and the radio communication range of a node is expressed as \textbf{\emph{r}}. These nodes can communicate only when ${\emph{\textbf{d}} \leq \textbf{\emph{r}}}$ .

Now we consider the four different cases in the velocities of these moving nodes and for each case link availability probability between nodes is discussed, these cases holds when ${\emph{\textbf{d}} \leq \textbf{\emph{r}}}$ :

\textbf{Case-1:} When both the nodes have same velocity and moving in same direction then link is available for longtime $t_1$ between them.

\textbf{Case-2:} In this case any of node has greater velocity than other one and moving in same direction then link is broken after some time $t_2$.

\textbf{Case-3:} This case deals that nodes have same velocity but moving in opposite direction then link is broken after some time $t_3$ but $t_3<t_2$.

\textbf{Case-4:} Here the case is different because both the nodes are moving in opposite direction with different velocities so link is broken after some time $t_4$ i.e: $t_4<<t_2$.

We discuss four cases above with velocities ${v_1,v_2 ~\exists~[V_{min}, V_{max}]}$ and ${\theta_{1,2}~ \exists~[0,\pi]}$ and now we will see the relative velocity between these nodes as:
\begin{eqnarray}
  \vec{{v_r}} &=& \vec{{v_1}}-\vec{{v_2}}
\end{eqnarray}
and using cosine law we can write it as
\begin{eqnarray}
  |\vec{{v_r}}| &=& \sqrt{{v_{1}}^2+{v_{2}}^2-2{v_{1}}{v_{2}}\cos{\theta_{1,2}}}
\end{eqnarray}
Hence relative velocity for four cases will be:

\textbf{Case-1:}  ${\vec{v_1}=\vec{v_2}}$ and angle ${:\theta_{1,2}=0}$ then ${|\vec{{v_r}}|=0}$.

\textbf{Case-2:}  ${\vec{v_1}=a\vec{v_2}}$, where $~\forall~a~\exists~(1,3)$ and angle ${\theta_{1,2}=0}$ then ${|\vec{{v_r}}|=(a-1)v_2}$.

\textbf{Case-3:} ${\vec {v_1}=\vec {v_2}}$ and angle ${\theta_{1,2}=\pi}$ then ${|\vec{{v_r}}|=2v_2}$.

\textbf{Case-4:}  ${\vec {v_1}=a{\vec {v_2}}}$, where $a~>~1$ and angle ${\theta_{1,2}=\pi}$ then ${|\vec{{v_r}}|=(a+1)v_2}$.

For the above cases a flow chart is given below in Fig. 1. This flow chart shows relative velocity $v_r$ for each of the case and also the link availability time $t_n$ between vehicles, where n=1 to 4.

\begin{figure*} [!t]
\begin{center}
\includegraphics[scale=0.55]{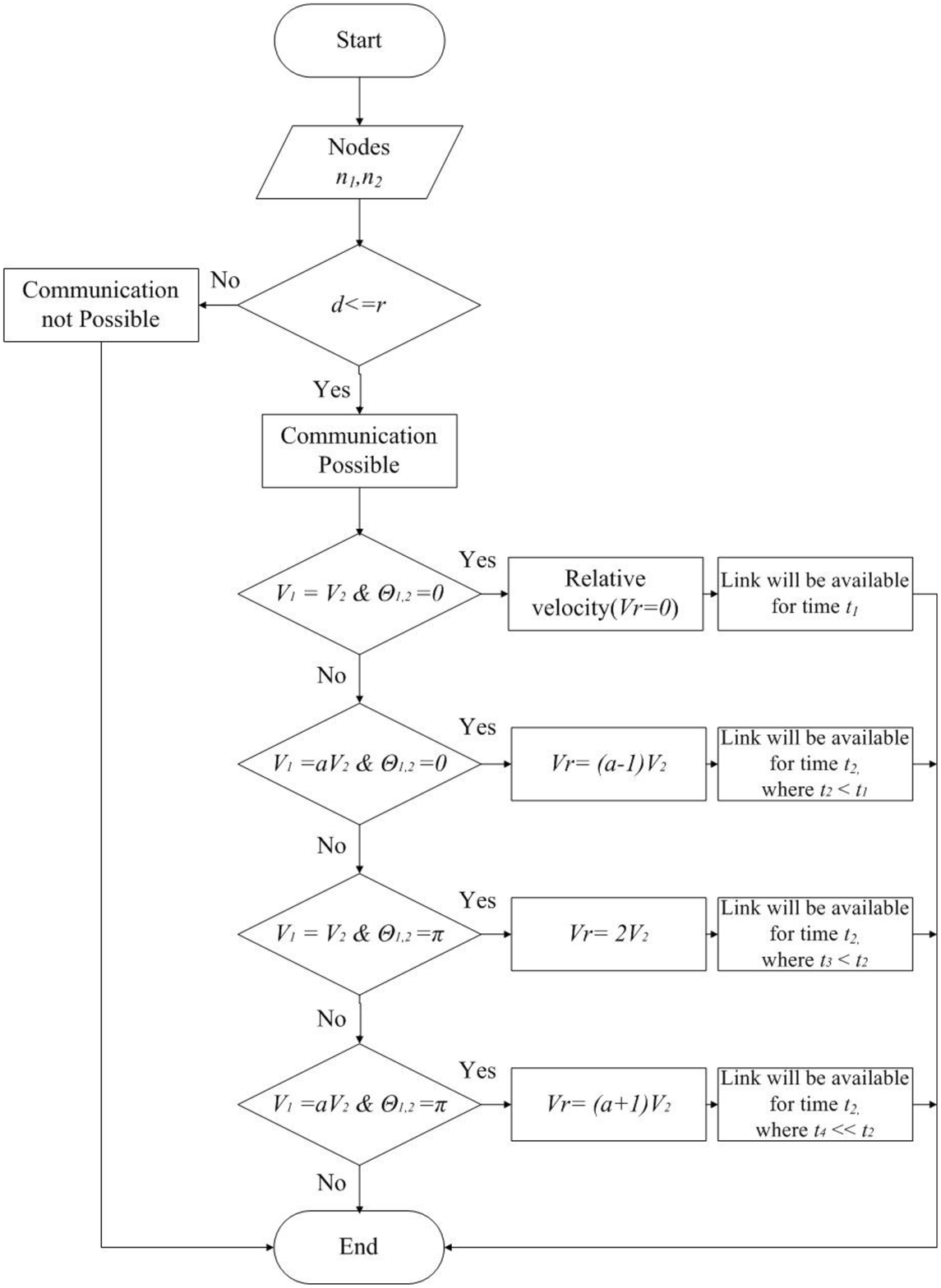}
\caption{Flow Chart of four cases of velocities}
\end{center}
\end{figure*}

From above results, it is observed that $v_r$ has different values so we represent it as random variable and according to probability density function (pdf), we can find it's expected relative velocity function as:

\begin{eqnarray}
  E(v_r) &=& \int^{\infty}_{-\infty} {v_r}f(v_r)\,dv_r
\end{eqnarray}

for further simplification ${f(v_r)=f(v_1)f(v_2)f(\theta_{12})}$, we can write eq. 3 as:

\small
\begin{eqnarray}
\centering
\begin{split}
E(v_r) = \int^{V_{max}}_{V_{min}} \int^{V_{max}}_{V_{min}} \int^{\pi}_{0} f(v_1)f(v_2)f(\theta_{12}) \\ {\sqrt{{v_{1}}^2+{v_{2}}^2-2{v_{1}}{v_{2}}\cos{\theta_{12}}}}\,d\theta_{12} \,dv_2 \,dv_1
\end{split}
\end{eqnarray}
\normalsize

Then expected relative velocity for each of the case discussed below.

\textbf{Case-1:} Nodes have same velocity and moving in same direction.

\begin{eqnarray}
\centering
E(v_r) = \int^{V_{max}}_{V_{min}} \int^{V_{max}}_{V_{min}} f(v_1)f(v_2) \,dv_2 \,dv_1
\end{eqnarray}

\textbf{Case-2:} Nodes have different velocity and moving in same direction.

\begin{eqnarray}
\centering
\begin{split}
E(v_r) = \int^{V_{max}}_{V_{min}} \int^{V_{max}}_{V_{min}} (a-1)v_2f(v_1)f(v_2) \,dv_2 \,dv_1 \\ ~\forall~a~\exists~(1,3)
\end{split}
\end{eqnarray}

\textbf{Case-3:} Nodes have same velocity and moving in opposite direction.

\begin{eqnarray}
\centering
E(v_r) = \int^{V_{max}}_{V_{min}} \int^{V_{max}}_{V_{min}} 2v_2f(v_1)f(v_2) \,dv_2 \,dv_1
\end{eqnarray}

\textbf{Case-4:} Nodes have different velocity and moving in opposite direction.

\begin{eqnarray}
\centering
\begin{split}
E(v_r) = \int^{V_{max}}_{V_{min}} \int^{V_{max}}_{V_{min}} (a+1)v_2f(v_1)f(v_2) \,dv_2 \,dv_1 \\ ~\forall~a~>~1
\end{split}
\end{eqnarray}

The expected relative velocity is function of random variable $v_r$, so this function shows random behavior for random values of $v_r$. Using this expected relative velocity function as random variable to find the probability of link availability, we develop an exponential probability density function given below:

\small
\begin{eqnarray}
  f(E(v_r)) &=& \frac{1}{dt_n}\exp(-\frac{E(v_r)}{dt_n})~~where~ n~=~1~to~4~
\end{eqnarray}
\normalsize

This pdf demonstrates the probability of available link between nodes. The function is exponential decaying function, so the term $\frac{E(v_r)}{dt_n}$ in eq. 9 increases, pdf decreases that means probability of link availability becomes small. Now we discuss for each of the case that how the pdf varies with varying expected relative velocity $v_r$ and link availability time $t_n$.

\textbf{Case-1:} From above explanation, it is observed that when nodes have expected relative velocity at $v_r = 0$ and the link availability time is $t_1$ that is large as compared to other cases. The term $\frac{E(v_r)}{dt_1}$ will have less value for large value of pdf and it illustrates that link availability probability for case-1 is greater than all the cases.

\textbf{Case-2:} In this case expected relative velocity has a value for $v_r$ between nodes is ${(a-1)v_2~\forall~a~\exists~(1,3)}$ and the link availability time is $t_2$ i.e: $t_2<t_1$. Now the term $\frac{E(v_r)}{dt_2}$ has more value than previous case so the probability for link availability is less than case-1.

\textbf{Case-3:} This case is different due to change in direction therefore, its expected relative velocity at $v_r = 2v_2$ is in eq. 7 and link availability time is $t_3$ i.e: $t_3<t_2$. Then link availability probability of nodes is smaller than case-2 because the term $\frac{E(v_r)}{dt_3}$ has large value due to increase in expected relative velocity and decrease in link available time $t_3$.

\textbf{Case-4:} In the last case the scenario is totaly different because in this case assumption is taken that nodes have different velocity and moving in opposite direction. Its expected relative velocity at $v_r=(a+1)v_2~\forall~a>1$ is in eq. 8 and the link availability time is $t_4$ i.e: $t_4<t_3$. Now this time the term $\frac{E(v_r)}{dt_4}$ is very very large because in this case $v_r$ has large value and $t_4$ has less value compare to all the cases. Therefore link availability probability between nodes will become so small from all the cases.

\begin{figure} [h]
\begin{center}
\includegraphics[scale=0.35]{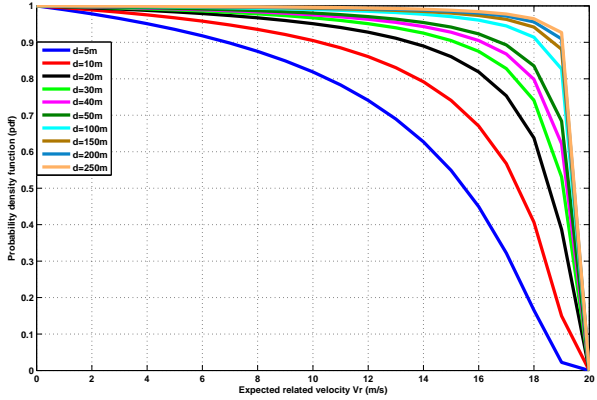}
\caption{Probability of link availability at random variable Vr}
\end{center}
\end{figure}

In order to analyze the cases that are discussed above, we create graph for eq. 9. From Fig. 2, we can say that link availability probability of two nodes decreases as expected related velocity increases. Another interesting thing is the distance $(\emph{\textbf{d}}<\emph{\textbf{r}})$ as it increases between nodes its graph shifts to upward due to increase in link availability probability.

\section{Experiments and Discussions }
In this paper, we use Nakagami propagation model in NS-2.34. The implementation of AODV and DSR used is the default one that comes with NS-2. For implementation of FSR, FSR patch is used [14]. The map imported in MOVE and scaled down to 4 km x 4 km in size for reasonable simulation environment. Using MOVE and SUMO, mobility patterns are generated randomly. Table I shows the complete simulation parameters used in simulations.

\begin{table}[h]
\caption{SIMULATION PARAMETERS}
\begin{center}
 \begin{tabular}{| c | c |}
  \hline
  \textbf{Parameters} & \textbf{Values} \\ \hline
    NS-2 Version &  2.34 \\ \hline
    AODV Implementation &	NS-2 default \\ \hline
    DSR Implementation	&   NS-2 default \\ \hline
    FSR Implementation  &	FSR-patch [8]\\ \hline
    MOVE version  &	  2.81 \\ \hline
    SUMO version  &	  0.12.3 \\ \hline
    Number of nodes & 20, 40, 60, 80, 100 \\ \hline
    Number of CBR sessions & 6, 12, 18, 24, 30, 36, 42 \\ \hline
    Tx Range &	300m \\ \hline
    Simulation Area &	4KM x 4KM \\ \hline
    Speed &	Uniform, 40kph \\ \hline
    Data Type &	CBR \\ \hline
    Data Packet Size &	1000 bytes \\ \hline
    MAC Protocol &	IEEE 802.11 Overhauled \\ \hline
    PHY Standard &	IEEE 802.11p \\ \hline
    Radio Propagation Model &	Nakagami \\ \hline

 \end{tabular}
\end{center}
\end{table}

\subsection{PDR}
Fig. 3 shows the average percentage of PDR against different number of connections and node density. In Fig. 3.(a), it is clear that AODV and DSR outperform FSR. AODV and DSR give almost same result and perform better than FSR.  The PDR of AODV increases as increase in the number of connections. Because it uses shortest path to destination and local link repair (LLR) mechanism, when link breakage occurs and consumes less bandwidth. The PDR of DSR also goes up as increment in the number of connections. Route reversal keeps away from the overhead of a possible second route discovery. FSR has low value of PDR due to proactive in nature because proactive routing protocols require more computation than reactive routing protocols. That is why, FSR shows average value of PDR as increment in number of connection.

\begin{figure}[h]
  \centering
\includegraphics[scale=0.35]{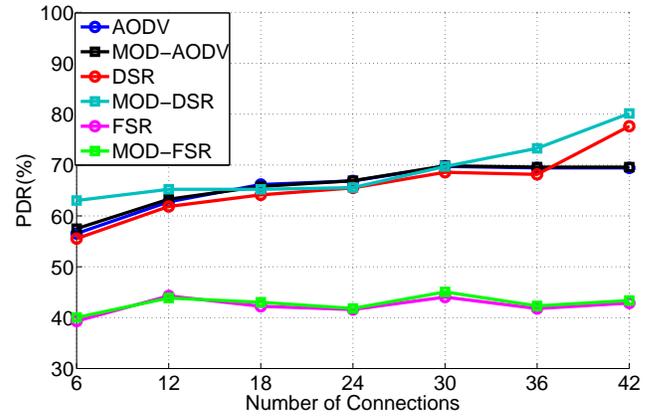}
 \caption{PDR vs No. of Connections}
\end{figure}

\begin{figure}[h]
  \centering
\includegraphics[scale=0.35]{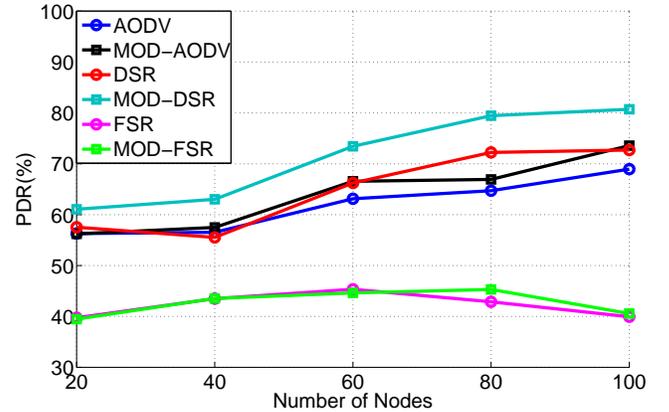}
 \caption{PDR vs Node Density}
\end{figure}

Fig. 3.(b) shows the PDR values of three routing protocols against node density. DSR gives slightly better PDR than AODV for high node density because valid routes are available in route cache, so it does not need more computation. Where as AODV uses the distance vector algorithm, so for every time using this algorithm, it finds the routes to destination. In node density, FSR performs same as for number of connections because it uses scope mechanism, due to this mechanism it shows average value for all the number of connections and node density.


From the results it is observed that link availability probability of DSR and AODV is higher than proactive routing protocol FSR because DSR has available routes in route cache, while AODV uses Hello messages for link sensing. Due to that reasons, their link availability time increases then the term $\frac{E(v_r)}{dt_n}$ have less value therefore eq. 9 results greater value of link availability probability. Whereas FSR has average value of PDR due to use of scope mechanism that causes less link availability time between nodes so its link availability probability is less than both reactive routing protocols.

\subsection{AE2ED}
Fig. 4 shows the AE2ED against different number of connections  and node density. From Fig. 4.(a), AE2ED of AODV is steady, but it is always more than both DSR and FSR. Due to increase in CBR sources, there is an increase in the number of packets contending for a common wireless channel, which leads to more collisions and more consumptions of bandwidth. So there is a significant drop in the delivery ratio and a corresponding increase in the AE2ED. The flow of AODV AE2ED that as number of connections increase, the delay decreases due to routing packets used in this protocol for establishment of path. The AE2ED flow of DSR is less than both routing protocols; AODV and FSR. But its delay initially decreases as connection increase (6 to 12) because it has available paths in route cache, if link breakage occurs, it checks route cache and uses another path. But its delay increases (18 to 42) as connections increase due to not getting available path in route cache, whenever link breakage occurs and broadcasts message RREQ for establishment of path. The AE2ED flow of FSR increase (12 to 24) as connections increase due to link breakage occurs because of dynamic movement in topology and source has to broadcast information to its neighbors, to spread the information to whole topology for establishment of path as a proactive routing protocol.

\begin{figure}[h]
  \centering
\includegraphics[scale=0.35]{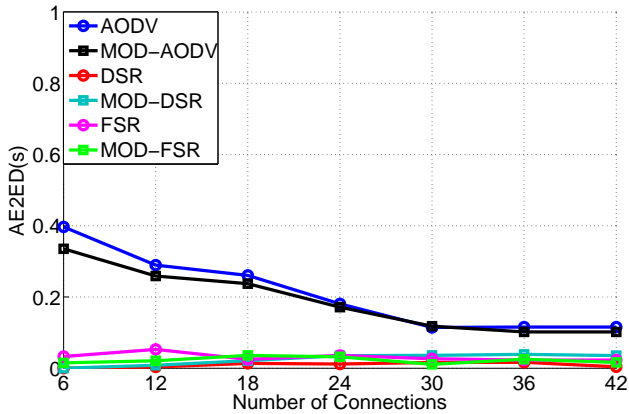}
 \caption{AE2ED vs No. of Connections}
\end{figure}

\begin{figure}[h]
  \centering
\includegraphics[scale=0.35]{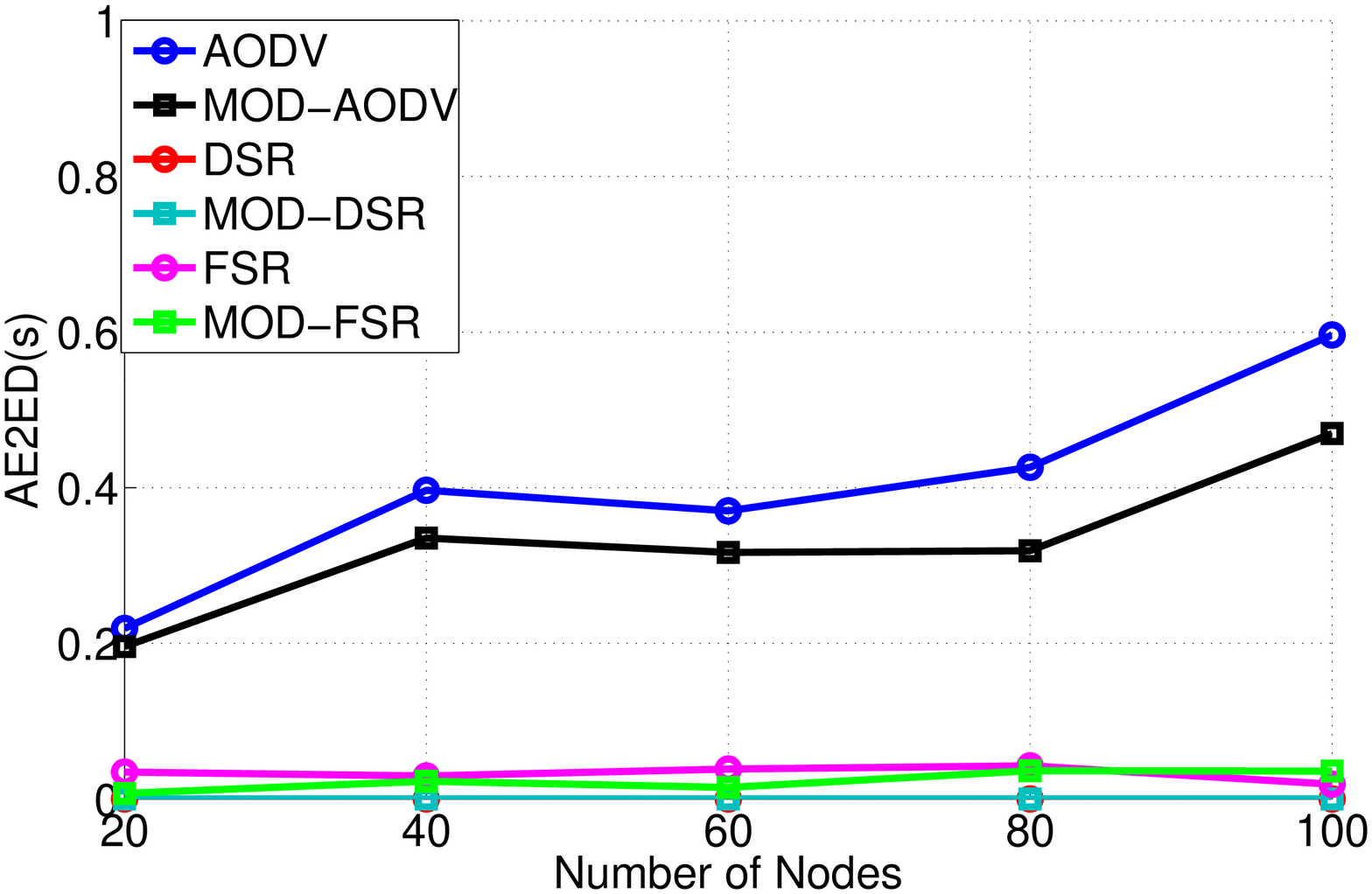}
 \caption{AE2ED vs Scalability}
\end{figure}

As it is clear from Fig. 4.(b), the AE2ED flow of AODV routing protocol is increasing as nodes increase due to two reasons. Firstly, the protocol uses LLR mechanism, to repair link breakage, which causes increase in path length. Secondly, the protocol uses message types, Route Requests (RREQs), Route Replies (RREPs), and Route Errors (RERRs) for establishment of Path which causes much delay. The flow decreases when there are less number of connections, because data packets transmit and then again for path establishment RREQ is broadcasted. AE2ED of DSR also increases due to message types but less than AODV's delay because it checks all available routes in Route cache whenever link breakage occurs. The AE2ED of FSR is more than DSR because, whenever link breakage occurs the source has to broadcast information to its neighbors to spread the routing information to whole topology. All paths are established, the data packets are transmitted to specified destination. Delay is marginally constant after path is established.

Modified routing protocols perform better than default routing protocols showing lowest value of AE2ED. MOD-AODV performs better than default one due to decrement in network Diameter. Whereas MOD-DSR shows good results than DSR due to decrement in cache size size. MOD-FSR has slightly better results than default FSR due to decreasing the interval of both inner and outer scope.


\subsection{NRO}
Fig. 5 shows the NRO of routing protocols against number of connections and node density. In Fig .5.(a), it is observed that NRO of FSR is less than both reactive routing protocols; AODV and DSR. As the number of connections increase, NRO of FSR decreases due to use of scope mechanism, that is good for more number of connection. FSR has lowest value among these routing protocols due to use of periodic updates to exchange topology map and also reducing the control messages. AODV shows the highest NRO value in the Fig. 5.(a) increasing graph for NRO. The reason for highest NRO of AODV is use of large number of control packets. DSR shows average behavior for NRO, as number of connections increase, due to stale entries in it's route cache.

In Fig. 5.(b), shows the NRO of routing protocols against node density. All the routing protocols show increasing NRO in high scalability. AODV has the same reason of its high NRO as mentioned above but for DSR in large number of nodes it generates Gratuitous Route Reply (grat. RREP) that causes more NRO. Whereas FSR has increasing graph due to scope mechanism.

Default routing protocols perform better than modified routing protocols producing less value of NRO. AODV performs better than MOD-AODV due to increment in TTL and threshold, and decrement in network size that cause better PDR and AE2ED but at the cost of highest NRO. Whereas DSR shows good results than MOD-DSR due to increment in buffer size and decrement in cache size causes more PDR and less AE2ED at the cost of large value of NRO. FSR has better results than MOD-FSR due to decreasing the interval of both inner and outer scope that causes better PDR and AE2ED than default FSR paying high value of NRO.


\vspace{-0.4cm}
\begin{figure}[h]
  \centering
\includegraphics[scale=0.35]{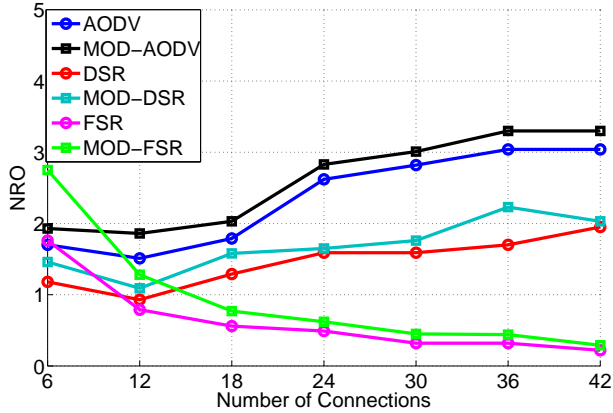}
\vspace{-0.4cm}
 \caption{NRO vs No. of Connections}
\end{figure}

\vspace{-0.4cm}
\begin{figure}[h]
  \centering
\includegraphics[scale=0.35]{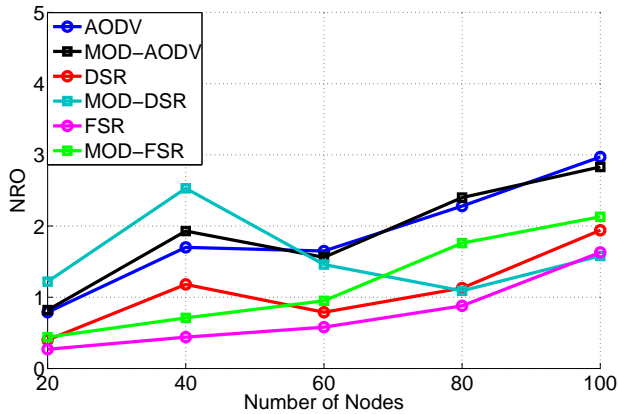}
\vspace{-0.4cm}
 \caption{NRO vs Scalability}
\end{figure}

\section{Performance Trade-offs Made by Routing Protocols}
From the results, we observe that each protocol performs best in any of the performance parameters such as: MOD-AODV and AODV sustain good PDR at some cost of AE2ED as number of connections and node density increase due to its LLR mechanism and distance vector routing. MOD-DSR DSR perform better in all three performance parameters; PDR, AE2ED and NRO with varying different number of connections and node density due to source routing and availability of routes in route cache. FSR produces less delay at the cost of low PDR in medium and high data flows and in all populations due to use of scope mechanism and link state routing.

\begin{table}[!h]
\caption {Efficiency (\%) Comparison of Default and Enhanced Protocols}
\begin {center}
\begin{tabular}{|c|c|c|}
\hline
\textbf{Results} & \textbf{Efficiency (\%) }&Efficiency (\%) \\
&\textbf{No. of Connections}&\textbf{No. of Nodes}\\\hline
$MOD\_AODV > AODV$&0.168&1.772\\\hline
$MOD\_DSR > DSR$&4.897&4.904\\\hline
$MOD\_FSR > FSR$&0.552&0.494\\\hline
$MOD\_AODV < AODV$&5.283&10.200\\\hline
$MOD\_DSR > DSR$&27.670&4.025\\\hline
$MOD\_FSR < FSR$&16.760&16.650\\\hline
$MOD\_AODV > AODV$&4.991&0.793\\\hline
$MOD\_DSR > DSR$&7.168&18.320\\\hline
$MOD\_FSR > FSR$&19.340&22.370\\\hline
\end{tabular}
\end{center}
\end{table}

\section{Conclusion}
In this paper, we evaluate and compare both default and modified routing protocols; AODV, MOD-AODV, DSR, MOD-DSR, FSR and MOD-FSR under the performance parameters; PDR, AE2ED and NRO. From the results, it is concluded that MOD-DSR and DSR perform better than routing protocols; AODV, MOD-AODV, FSR and MOD-FSR, in terms of PDR and AE2ED at the cost of high value of NRO. Whereas, AODV performs better for PDR and NRO and sustain high value of AE2ED with varying number of connections and node density. MOD-FSR and FSR show less AE2ED and NRO at the cost of less value of PDR. MOD-DSR and DSR outperform AODV, MOD-AODV, OLSR and MOD-OLSR in terms of link availability probability at the cost of high value of NRO.

\end{document}